\documentclass[twocolumn,twocolappendix]{aastex6}

\newcommand\Mp{M_{\rm p}}

\newcommand\Mhi{M_{\rm HI}}
\newcommand\Mtot{M_{\rm tot}}

\newcommand{\Msun}{M_{\odot}}
\newcommand\kpc{{\rm \,kpc}}
\newcommand\pc{{\rm \,pc}}
\newcommand\Gyr{{\rm \,Gyr}}
\newcommand\Myr{{\rm \,Myr}}
\newcommand{\Rstar}{R_{\ast}}
\newcommand{\vx}{v_{x}}
\newcommand{\vy}{v_{y}}
\newcommand{\vz}{v_{z}}
\newcommand\kms{{\rm \,km\,s^{-1}}}
\newcommand\degree{^{\circ}}

\begin{document}


\title{DDO~68: A  flea with smaller fleas that on him prey}


\author{Francesca Annibali\altaffilmark{1,2} }
\altaffiltext{1}{INAF - Bologna Observatory, via Ranzani 1, 40127 Bologna, Italy}
\author{Carlo Nipoti\altaffilmark{2}}
\altaffiltext{2}{Department of Physics and Astronomy, University of Bologna, viale Berti-Pichat 6/2, 40127 Bologna, Italy}
\author{Luca Ciotti\altaffilmark{2}}
\author{Monica Tosi\altaffilmark{1} }
\author{Alessandra Aloisi\altaffilmark{3}}
\altaffiltext{3}{Space Telescope Science Institute, 3700 San Martin Drive\,Baltimore MD 21218}
\author{Michele Bellazzini\altaffilmark{1}}
\author{Michele Cignoni\altaffilmark{4}}
\altaffiltext{4}{Physics Department, University of Pisa, Largo Bruno Pontecorvo 3, 56127 Pisa, Italy}
\author{Felice Cusano\altaffilmark{1}}
\author{Diego Paris\altaffilmark{5}}
\altaffiltext{5}{INAF-Rome Observatory, Via Frascati 33, 00078 Monte Porzio, Italy}
\and
\author{Elena Sacchi\altaffilmark{1,2}}

\begin{abstract}
We present new photometry of the dwarf irregular galaxy DDO~68, one of  
the most metal-poor and least massive dwarfs, located in the Lynx-Cancer Void. 
The images were acquired with the Large
Binocular Telescope in the g and r passbands, and show unequivocally
that DDO~68 has previously unknown stellar streams related to the accretion of at least
two smaller companions: {\it a flea with smaller fleas biting it}, to put it
in Jonathan Swift's words\footnote{From Jonathan Swift$'$s {\it On Poetry: a Rhapsody}: So, naturalists observe, a flea/ has smaller fleas that on him prey;/ and these have smaller still to bite Ôem/ and so proceed ad infinitum.}.  Our data provide direct observational
evidence of multiple merging occurring at very low galactic mass
scales. We present the results of an $N$-body simulation of the interaction of three
dwarf galaxies which reproduces well the main morphological features
of DDO~68.
\end{abstract}




\section{Introduction} \label{sec:intro}

In Lambda Cold Dark Matter cosmology, galaxies grow continuously in mass through
hierarchical assembly of smaller systems \citep[e.g.][]{White78,Diemand08}, and dark matter 
haloes host substructures down to the
resolution limit of the simulations \citep{Diemand08,Wheeler15}.
Observationally, there is ample evidence of accretion of satellites
onto massive galaxies such as the Milky Way, Andromeda, several Local
Volume spirals, and the giant elliptical Centaurus A \citep[see,
  e.g.,][]{Ibata01,Belokurov06,McConnachie09, Martinez-Delgado10,
 Crnojevic16}. Yet, evidence of stellar streams around dwarf galaxies
is currently limited \citep{Rich12, Martinez-Delgado12, Belokurov16}; in particular, no clear cut stellar stream 
has been detected so far around galaxies less massive in stars than $10^9 M_{\sun}$,  
 except for a kinematical signature suggesting that the dwarf satellite of M31 - And~II -
ingested a smaller system in the past \citep{Amorisco14}.

DDO~68 is a star-forming dwarf galaxy located in the Lynx-Cancer void
\citep{Pustilnik11} at a distance of $\simeq$12.7 Mpc from us
\citep{Cannon14, Sacchi16}. With an oxygen abundance \citep{Pustilnik05,Izotov09} about 1/40 of
solar \citep{Caffau08}, it is one of the most
metal-poor star-forming dwarfs known so far, much more metal-poor than
typical dwarfs of similar mass \citep{Pustilnik05}. It also exhibits a
very distorted morphology, with a cometary tail (hereafter, the Tail)
populated by stars of all ages \citep{Tikhonov14, Sacchi16} and particularly rich in H~II regions, 
that extends from the main body of the galaxy for a projected
length of $\simeq$5 kpc. Because of these peculiar
properties, DDO~68 has already been suggested to be affected by galaxy
interaction \citep{Ekta08,Cannon14,Tikhonov14}, although with no
conclusive evidence. Ekta et al. noticed distortions in the H~I
distribution and interpreted the observed features in terms of a late-stage merger of two gas-rich progenitors. 
Tikhonov et al. proposed that the Tail is actually a
disrupted satellite (dubbed DDO~68~B) being currently
accreted by DDO~68, while Cannon et al. identified a possible H~I
satellite (DDO~68~C), with $\Mhi\simeq3 \times 10^7 \Msun$, at a projected distance of $\simeq$40 kpc from the main body.

\begin{figure*}[ht!]
\plotone{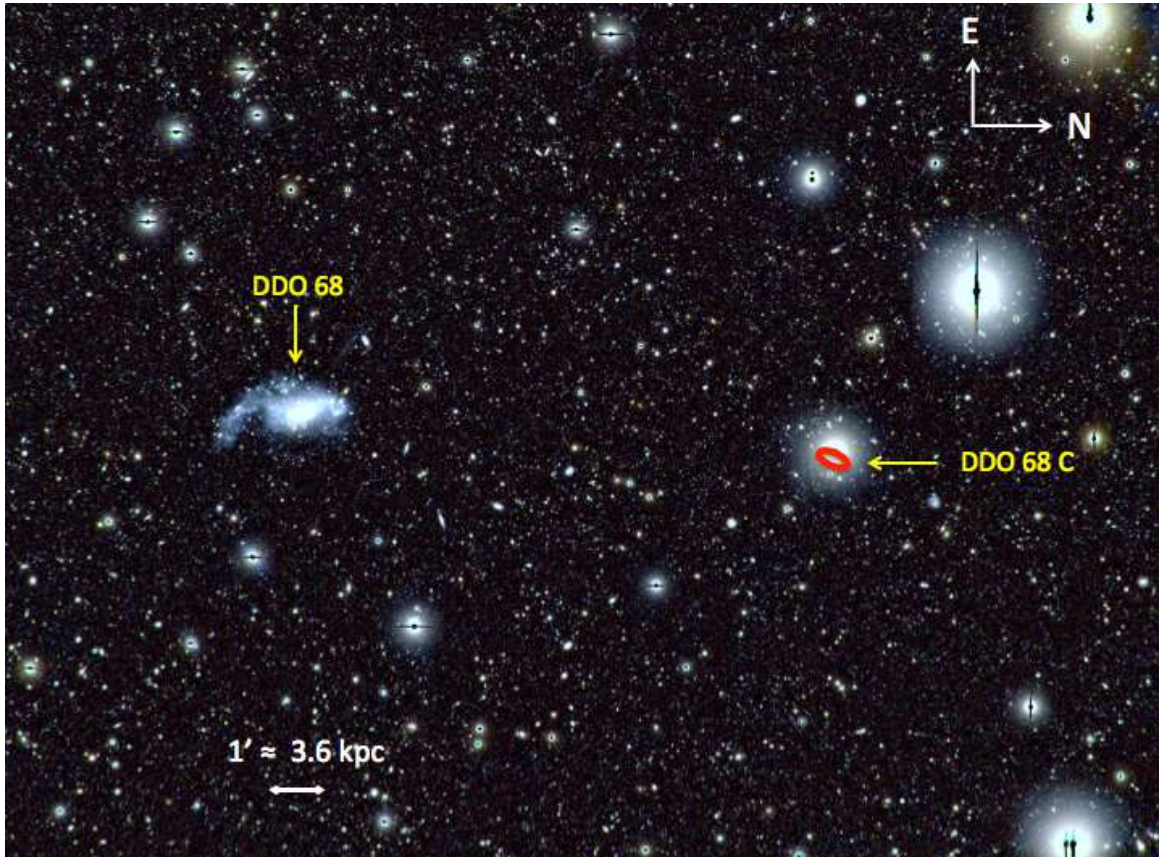}
\caption{Deep LBC g and r color-combined image of DDO~68. 
 The positions of DDO~68 and of
  the candidate H~I companion DDO~68~C are indicated. DDO~68~C is 
  masked by a bright foreground star, but we have
  drawn its spatial extension (red ellipse) as detected in the
  ultraviolet GALEX images. The total displayed field of view is $\simeq
  0.4 \times 0.3$ square degrees, corresponding to $\simeq 90 \times 70$
  kpc$^2$ at DDO~68's distance of $\simeq$12.7 Mpc.
\label{lbtimage} }
\end{figure*}

We studied DDO~68 with deep images acquired with the Advanced Camera
for Surveys (ACS) on board of the Hubble Space Telescope (GO
11578, PI Aloisi). With those data we \citep{Sacchi16}
resolved its stellar content, characterized its star formation history
(SFH) with the synthetic colour-magnitude diagram (CMD) method
\citep{Tolstoy09,Cignoni10}, and constrained its distance via the tip
of the red giant branch (RGB) and the synthetic CMDs. From the
resulting SFH we inferred that the total mass in stars ever formed in
the system is only $1.2 \times 10^8 M_{\sun}$. DDO~68 is of special
interest as a possible accretor, because of this low mass, that puts
it close to the resolution limit of galaxy formation models.

The ACS field of view is too small to look for
satellites or streams around the galaxy. To obtain
very deep imaging that could reveal the presence of previously unknown 
faint stellar substructures connected to DDO~68, we exploited the perfect
combination of large field of view and excellent photometric depth of
the LBT Large Binocular Camera (LBC) covering a large area around our
target. In this Letter we report the successful results obtained from
the LBT photometry, with clear evidence that, in addition to the
Tail and DDO~68~C, the system hosts also an independent small stream, 
and several arc-like structures. To better interpret the LBT data in terms of the
dynamical status of the system we have also run $N$-body simulations,
and found that the observed configuration is consistent with the
presence of at least two different satellites being accreted by the
galaxy main body.

\begin{figure*}[ht!]
\plotone{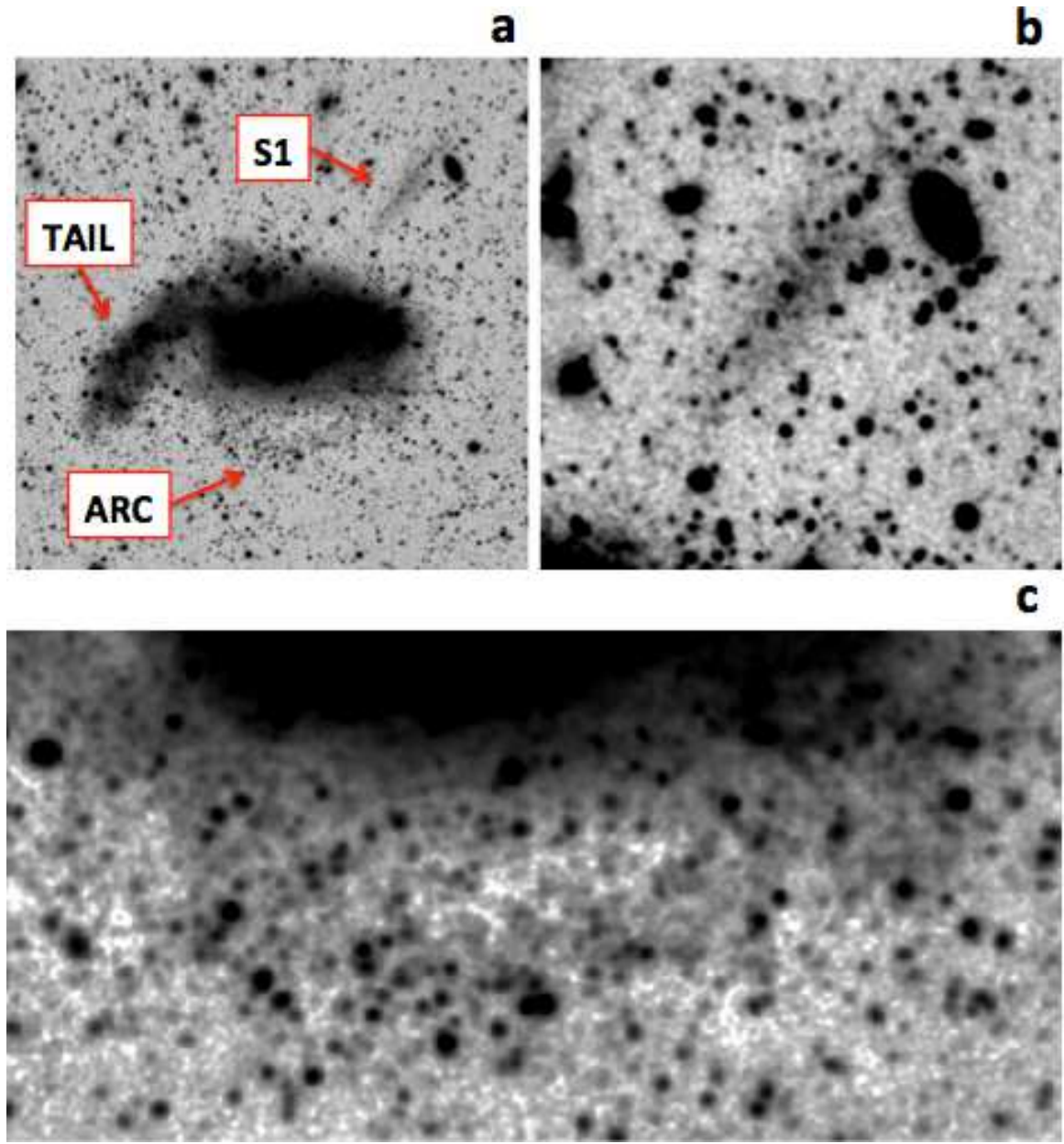}
\caption{Blow-ups of key portions of the LBT image of DDO~68. {\bf a}:
  Portion of the g-band image (FoV $\simeq280 \times 280$ arcsec$^2$, or
  17 $\times 17$ kpc$^2$) showing the Tail, Arc, and stream
  S1. A Gaussian smoothing has been applied to the image to better
  highlight low-surface brightness features. {\bf b}: Further blow-up
  (FoV $\simeq90 \times 90$ arcsec$^2$, or 5 $\times$ 5 kpc$^2$) in the
  r band showing S1 in detail. {\bf c}: Blow-up of panel {\bf a} (FoV
  $\simeq140 \times 60$ arcsec$^2$, or 8.4 $\times$ 3.6 kpc$^2$) in the
  g band showing the Arc in detail.
\label{lbt-blowup} }
\end{figure*}

\section{LBT photometry}

We observed DDO~68 with LBT in January 12 and 13, 2016, with the wide
field ($\simeq$23 arcmin $\times$ 23 arcmin) LBC cameras operated in
binocular mode. Images in the Sloan Digital Sky Survey (SDSS) g and r
passbands were obtained with the LBC-blue and LBC-red camera,
respectively. The total exposure time of $\simeq$2 h in each band was
organized into 5 visits of 180 s times 8 dithered exposures. For a
straightforward calibration into the SDSS photometric systems, we also
acquired in each band 4 short exposures of 30 s each to serve for the
photometry of bright foreground stars in common with the SDSS
catalog. With our exposures, we reached a
surface-brightness limit of $\mu\simeq29$ mag arcsec$^{-2 }$ in the g
and r bands.

The reduction of the g- and r- LBC images was performed with the
pipeline developed at INAF-OAR (Paris et al. in
preparation). The individual raw images were first corrected for bias
and flat field, and then background-subtracted. After astrometric
calibration, they were combined into a g- and r-band stacked images
with the SWarp software \citep{Bertin02}.

Our color-combined LBT image is displayed in Figure~\ref{lbtimage},
where both DDO~68 and the position of the candidate companion DDO~68~C
\citep{Cannon14}
are indicated. The new substructures/satellites discovered here are
much closer to the main body of DDO~68, and are visible in 
Figure~\ref{lbt-blowup}, where blow-ups of some portions of the whole image
are displayed.
Remarkably, Figure~2, besides the already well known Tail,
reveals several previously undetected faint stellar substructures.
The most prominent of these are an arc embracing the
western side of DDO~68 for a projected extension of $\simeq$5 kpc
(hereafter, the Arc) and a stream (which we dub DDO~68-S1, i.e. Stream
1) as faint as $\mu_r\simeq$28.7 mag arcsec$^{-2}$ that extends for
$\simeq$4.6 kpc along the NE-SW direction (the system at the NE end of
S1 is most likely a background spiral galaxy).
While S1 is well detected in both the g and r images, the Arc is
mostly visible in g, due to the presence of younger stars than in S1 (see Section~3).

\begin{figure*}[ht!]
\plotone{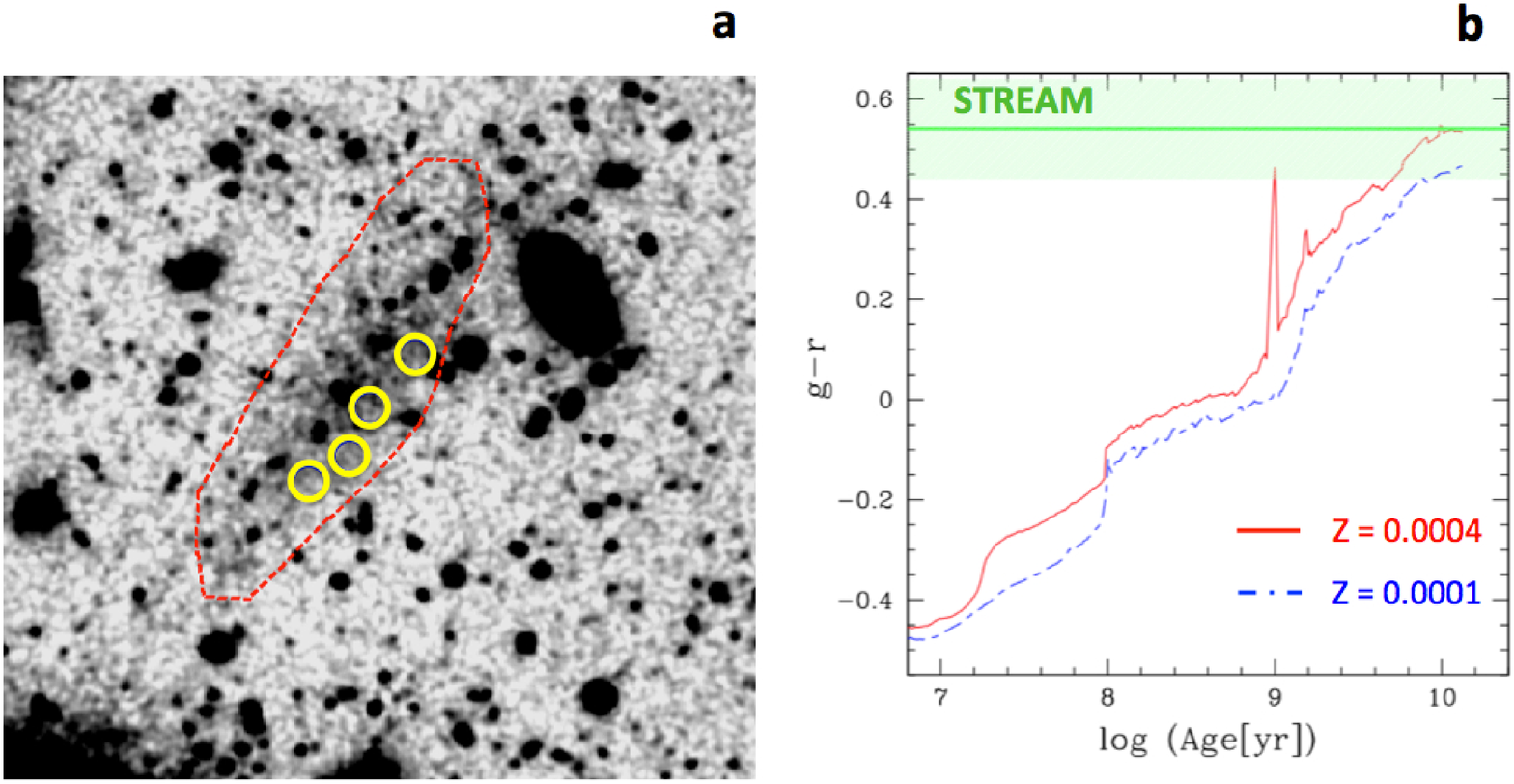}
\caption{{\bf a}, LBT r-image of the stream  DDO~68-S1 (FoV $\simeq90 \times 90$ arcsec$^2$) with
  overlaid both the region (red dashed line) selected to compute its total magnitude, 
  and the circular apertures,
  free of visible contaminating galaxies, selected to compute its color. {\bf b}: g-r color of SSP models as a
  function of age. The plotted models are the PARSEC ones 
  for the two metallicities Z=0.0001, 0.0004, and a Kroupa IMF.
  The green horizontal line and the shaded area
  indicate the intrinsic (g-r)$_0$ color of S1 and its $\pm1\sigma$
  uncertainty. \label{s1} }
\end{figure*}

The total magnitude of S1 in the r band was derived within the region
drawn in Figure~\ref{s1} using the {\it polyphot} task in the {\it
  apphot} IRAF\footnote{IRAF is distributed by the National Optical Astronomy Observatory, operated by AURA} 
 package. The background was evaluated drawing the same area in regions randomly
placed around S1  avoiding both DDO~68 and bright
foreground stars, and performing photometry in the same way as for
S1. This procedure allowed us to remove the contribution from both the
sky and the background galaxies. The average (g-r) color was instead
computed performing photometry within a few small (2 arcsec in radius)
circular apertures selected in regions of S1 free of visible
contaminants (see Figure~\ref{s1}a), with the background evaluated in
similar apertures placed in free regions around S1.  Finally, the
calibration of the instrumental magnitudes into the SDSS system was
accomplished through photometry on our short-exposures of stars
catalogued in the SDSS.

We obtain a total integrated magnitude for S1 of m$_{r,S1} = 21.6 \pm
0.4$ and average color of (g-r)$_{S1} = 0.56 \pm 0.11$, computed
within an area of $\simeq$700 arcsec$^2$. This provides an average
surface brightness $\mu_{r,S1}\simeq28.7$ mag arcsec$^{-2}$.  Adopting
a distance of 12.7 Mpc \citep{Sacchi16}, or (m-M)$_0
\simeq$30.5, and a foreground reddening of E(B-V) = 0.018
\citep{Schlegel98}, we derive an absolute intrinsic magnitude
M$_{r,S1} = -9.0 \pm$ 0.4, and an intrinsic color (g-r)$_{0,S1} = 0.54
\pm$0.11.

\begin{figure*}[ht!]
\plotone{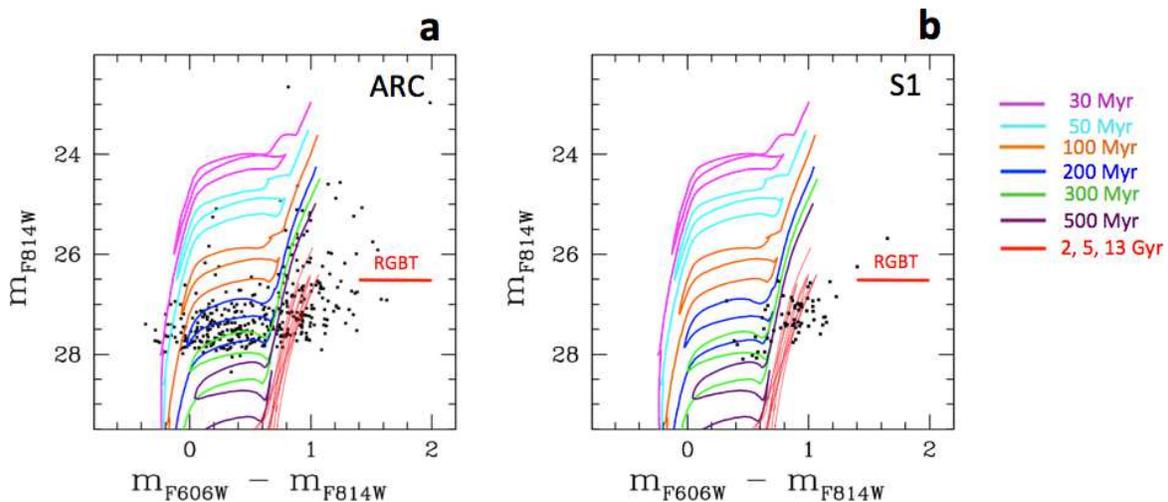}
\caption{HST color-magnitude diagrams of the Arc and of S1 in DDO~68. {\bf a}: CMD of the stars resolved in our ACS@HST images
  that belong to the region outlined by the Arc. Superimposed are the
  PARSEC isochrones \citep{Bressan12} for a metallicity Z=0.0004
  and for different ages, shifted according to DDO~68's distance and
  foreground reddening. The horizontal segment indicates the
  location of the tip of the red giant branch as derived by
  \cite{Sacchi16}. {\bf b}: same as {\bf a}, but for the small portion
  of S1 falling within the HST images. S1 is mainly populated by RGB
  stars, implying ages older than 2 Gyr. \label{cmd} }
\end{figure*}

\section{Physical properties of the substructures}

Both the Arc and a small portion of S1 fall into our ACS@HST images 
and are resolved into individual stars.  Their CMDs (see Figure~\ref{cmd}) host 
RGB stars  with luminosities compatible with DDO~68's distance. This indicates that both structures are physically
associated with DDO~68.

The CMD of the Arc implies a wide range of ages, with most of the
measured stars $\simeq$200-300 Myr old, a few as young as 50 Myr,
and several RGB stars older than 2 Gyr and possibly up to 13
Gyr old. Unfortunately, the age-metallicity degeneracy of the RGB colors,
coupled with the large photometric errors, prevents us to safely
distinguish a 2 Gyr old from a 13 Gyr old population in DDO~68. 

The resolved portion of S1 exhibits a population older than the Arc,
since its CMD is only populated by RGB stars, with ages $\gtrsim$2 Gyr.  To further characterize S1's stellar population, we
compared its intrinsic color derived in the previous Section with the 
PARSEC v1.2S + COLIBRI PR16 simple stellar population (SSP) models  
\citep{Bressan12,Marigo13,Rosenfield13},  for a Kroupa initial mass function \citep[IMF, ][]{Kroupa01},
at DDO~68' s metallicity of Z=0.0004
($\simeq$1/40 of solar, where Z$_{\sun}$ = 0.0152) and at a lower
metallicity of Z=0.0001 ($\simeq$1/150 solar). The S1 color implies,
within the 1$\sigma$ error, a population older than 5 Gyr, and
possibly as old as the age of the Universe (Figure~\ref{s1}b). This is in agreement with the fact that the portion of S1
sampled by our HST data is resolved into RGB stars (Figure~\ref{cmd}),
which can have ages in the range 2-13 Gyr. From the SSP models, we
derive a total S1 luminosity of $\simeq3.5\times10^5$ L$_{\sun}$. To derive
a stellar mass range for S1, we conservatively adopted two age
extremes at 2 Gyr and 13 Gyr, which provide a current stellar mass in
the range $(1.5-6) \times 10^5 \Msun$. S1 is then comparable to the
Ultra Faint satellites of the Milky Way
\citep{Belokurov07,Belokurov10}, although it may also be just a
portion of a larger disrupted progenitor.

The significantly different stellar populations of the Arc and S1
suggest that the two substructures originated from two different
systems.

\section{Dynamics of DDO~68}

\begin{figure*}
\plotone{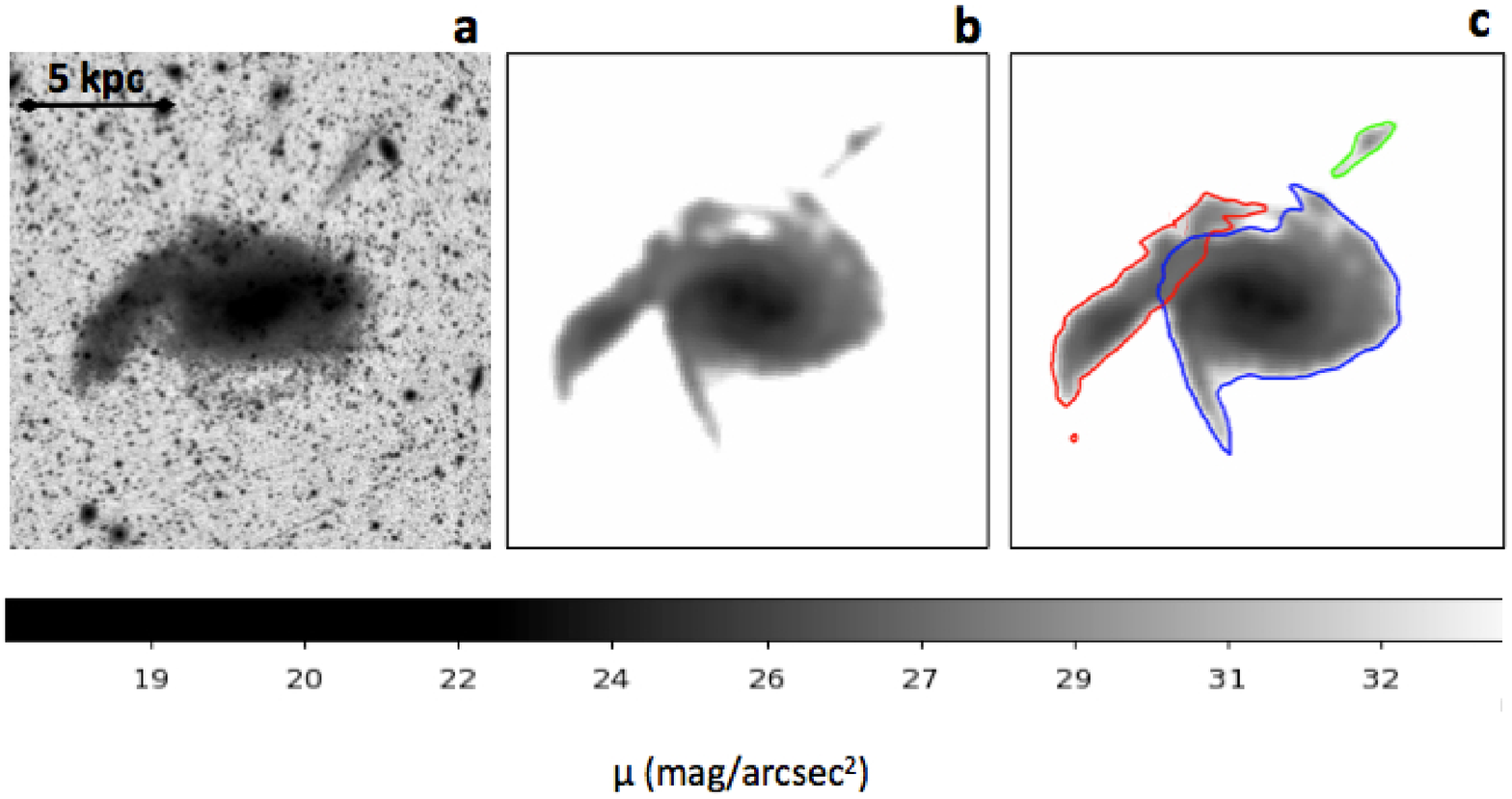}
\caption{{\bf a}: g-band image of DDO~68 in surface brightness
  scale. {\bf b}: snapshot of the stellar component of the
  multiple-merger $N$-body simulation. The observed and the simulated
  images are displayed on the same spatial and surface-brightness
  scales. We note that elongated arc-like substructures, akin to the
  observed Arc (see Fig. 2a), are present in the mock image.  {\bf c}:
  final snapshot of the $N$-body simulation with superimposed the
  contours delimiting the regions whose bulk stars originally belong
  to the main galaxy (blue), to satellite T (red) and to satellite S
  (green).  
 \label{sim} }
\end{figure*}

A possible interpretation of the morphology of DDO~68 is that the
Tail, the Arc and S1 are tidal features due to the stripping effect of
a passing body. However, no nearby objects in the direction of the
Tail, the Arc, or S1 are detected in Fig. 1 or in the Sloan Digital
Sky Survey images. A simple estimate of the mass $\Mp$ of a putative
perturber at distance $d$ from the main body of DDO~68 is given 
by the fact that a significant stripping can be produced only for 
$\Mp\simeq \Mtot d^3/R^3$, where $\Mtot$ is the total mass and $R$ is
a fiducial radius of DDO~68. This leads to deduce implausibly high
values for the mass of an otherwise undetected 
perturber.  For example, the minimum mass of a
perturber located at a distance $d$ three times the length of the Tail
($R\simeq5\kpc$) would be $\Mp \simeq 30 \Mtot$.  
This argument excludes that the Tail, the Arc and S1 are tidal features due to DDO~68~C, 
which is at a distance of 42 kpc \citep{Cannon14} from the main body, 
and has a baryonic mass $\approx$10 times smaller than that of DDO~68.

Being tidal stripping excluded as the cause of the three
non-equilibrium features, the alternative is that
they are due to the accretion of smaller companions. Support to the 
interpretation of DDO~68 as an ongoing multiple merger comes from the
results of $N$-body simulations. We ran several simulations
of possible merging events in DDO~68 with the parallel
collisionless $N$-body code FVFPS \citep{Nipoti03}. 
For simplicity, we include  only the collisionless component 
(stars and dark matter) in the simulations, and we do not model the physics of the gas. 
We found that while the Tail and a structure similar to
the Arc are reproduced by the accretion of a ten times less-massive
satellite galaxy (with the Tail surviving for $\simeq 100 \Myr$ and
consisting mainly of stars of the satellite), the formation of S1 requires an
additional accreting system. We thus performed simulations (see Appendix~\ref{app:nbody} for details)  
in which the main galaxy of mass $\Mtot$ interacts simultaneously with two
smaller satellites, named T and S, of mass $\Mtot/10$ and
$\Mtot/150$, respectively.  As an example we
compare in Fig.~\ref{sim} the LBT image of DDO~68 with a snapshot of one of
these $N$-body simulations  which reproduces well all its 
main morphological features.  In this
snapshot, the spatial distributions of the stars of satellites T and S
correspond, respectively, to the Tail and to S1. 

\section{Conclusions}

Our study of DDO~68 shows clear evidence of stellar streams
around a galaxy with stellar mass as low as M$_{star} \simeq10^8$ M$_{\sun}$, and located
in a void. DDO~68 is an extremely isolated system and yet it appears
surrounded by smaller, interacting bodies: the Tail \citep[or DDO~68~B, in the notation of][]{Tikhonov14}, DDO~68~C \cite[discovered by][]{Cannon14}, the Arc and S1, 
detected now with LBT.

A simple dynamical analysis suggests that the observed morphological features (Tail, Arc and S1) are not 
the effect of tidal interaction with massive companions, but are instead the signature of an ongoing multiple merging, 
as confirmed by our N-body simulations. The simulation, which is admittedly simplified, must be considered only an 
illustrative case and not necessarily the best possible way to reproduce the complex morphology of DDO~68. 

Multiple accretion of dwarf
systems onto a more massive host is expected from observations showing
that dwarfs are often found in associations \citep[see,
  e.g.][]{Tully06,Bellazzini13} and from cosmological simulations
predicting that sub-halos are often accreted in small groups
\citep{Li08} - a scenario that could explain the association of some
Milky Way dwarfs with the plane of the orbit of the Magellanic Clouds
\citep{Donghia08}. In DDO~68, we are witnessing multiple accretion of
smaller systems in action. We suggest that it is not just a
coincidence that these {\it fleas of fleas} are so clearly observed for
the first time in such an isolated system: once a group of dwarfs with
their satellites falls into a massive host, like the Milky Way, it is
eventually disassembled by tidal forces wiping out evidence of
coherent structure \citep{Deason15,Wheeler15}. It is natural to expect
satellites of satellites residing in small haloes, i.e. in lower
density environments, to survive longer than those residing in more
massive haloes.

While all dwarf irregulars show distorted morphologies, by definition, this does
not imply that these are produced by recent merging events. Here we have identified
very specific signatures  (like, e.g., a stream made of old stars, that cannot
originate from internal  hydrodynamical processes) that are not usually seen in
other dIrrs. Yet, lack of evidence is not evidence of lack, and might instead be due
to insufficient sensitivity of the available observations. Our result 
demonstrates the potential of wide-field instrumentation at 8-10 m telescopes to
search for substructures around dwarfs. 

\acknowledgements We thank V. Belokurov for the nice review. This letter is based on LBT images acquired within
the Italian Director Discretionary Time, for which we are grateful. 
We acknowledge support from the LBT-Italian Coordination Facility and the
Italian LBT Spectroscopic Reduction Center for the execution of
observations, data distribution, and reduction. 
F.A. was partially
funded through the Italian PRIN-MIUR 2010LY5N2T-006.

\appendix

\section{$N$-body simulation}
\label{app:nbody}

In the initial conditions of the $N$-body simulation shown in
Fig.~\ref{sim}b the main galaxy is represented by an equilibrium
spherical isotropic dark-matter halo and a stellar disc. The halo has
$\gamma=0$ density profile \citep{Dehnen93} $\rho(r)\propto(r+a)^{-4}$
with scale radius $a=4\kpc$ and total mass $1.3\times10^{10}\Msun$
(the mass within $11\kpc$ is $5.3\times10^9\Msun$, consistent with the
observational estimate of the dynamical mass of DDO~68;
\citealt{Cannon14}). The stellar disc is assumed to be razor-thin, with exponential surface density profile
$\Sigma(R)\propto\exp(-R/\Rstar)$ with $\Rstar=1.02\kpc$.
This profile provides a satisfactory fit to the completeness-corrected stellar counts as a function of radius 
as derived from the HST photometry.
In the simulation, the stellar particles are in circular orbits in the gravitational potential of the
halo. The gravitational potential of the stars is neglected, which is
justified because the dynamical mass is about 20 times the stellar mass within a
radius of $11\kpc$ for DDO~68.

The satellites are scaled-down versions of the same model used for the
main galaxy (spherical $\gamma=0$ halo and thin exponential stellar
disc). Satellite T has halo scale radius $a=1.86\kpc$, total mass
$1.3\times10^9\Msun$, and stellar scale radius
$\Rstar=0.47$. Satellite S has halo scale radius $a=1.2\kpc$, total
mass $8.6\times10^7\Msun$, and stellar scale radius
$\Rstar=0.3\kpc$. As for the main galaxy, also the stellar particles
of the satellites are just tracers: their gravitational potential is
neglected. The number of dark-matter particles is $3\times10^5$ for
the main galaxy, $3\times10^4$ for satellite T, and $2\times10^3$ for
satellite S. In each system the stellar disc is represented with the
same number of particles used for the dark-matter halo.

Taking a Cartesian coordinate system with origin in the centre of mass
of the main galaxy, the initial ($t=0$) centre-of-mass positions and
velocities are $x=10\kpc$, $\vy=-45\kms$, $y=z=\vx=\vz=0$ for
satellite T, and $x=-10\kpc$, $y=4.37\kpc$, $\vz=45\kms$,
$z=\vx=\vy=0$ for satellite S (both satellites are approximately in
circular orbit in the gravitational potential of the main galaxy). The
orientation of this Cartesian system is such that the $y$ axis lies in
the plane of the main-galaxy disc, which forms an angle of $45\degree$
with the $x$ axis. The disc of satellite T lies in the satellite's
orbital plane, while the disc of satellite S is orthogonal to the
satellite's orbital plane. In the simulation the evolution of the
system is followed for $1\Gyr$. The time step, which is allowed to
vary with time as a function of the maximum mass density
\citep{Nipoti03}, is typically in the range $1-2\Myr$. The softening
length is $30\pc$. The snapshot shown in Fig.~\ref{sim}b is obtained
at $t = 630\Myr$ assuming a line of sight such that the disc of the
main galaxy at $t=0$ has an inclination of $60\degree$.



\end{document}